# Enhancement of the transverse non-reciprocal magneto-optical effect

V. Zayets, H. Saito, S. Yuasa and K. Ando

*Spintronic Research Center, National Institute of Advanced Industrial Science and Technology (AIST), Umezono 1-1-1, Tsukuba, Ibaraki, Japan. E-mail: v.zayets@aist.go.jp*

The origin and properties of the transverse non-reciprocal magneto-optical (nMO) effect were studied. The transverse nMO effect occurs in the case when light propagates perpendicularly to the magnetic field. It was demonstrated that light can experience the transverse nMO effect only when it propagates in the vicinity of a boundary between two materials and the optical field at least in one material is evanescent. The transverse nMO effect is pronounced in the cases of surface plasmons and waveguiding modes. The magnitude of the transverse nMO effect is comparable to or greater than the magnitude of the longitudinal nMO effect. In the case of surface plasmons propagating at a boundary between the transition metal and the dielectric it is possible to magnify the transverse nMO effect and the magneto-optical figure-of-merit may increase from a few percents to above 100%. The scalar dispersion relation, which describes the transverse MO effect in cases of waveguide modes and surface plasmons propagating in a multilayer MO slab, was derived.

The magneto-optical (MO) effect is important for a variety of applications. The effect is utilized to read data in a MO disk driver[1], to switch an optical beam in optical switches[2] and to modulate light intensity in spacial light modulators[3]. It is a powerful scientific tool to determine the local magnetization of a material, to study the bandgap structure and spin-orbit interaction in a solid[4]. A unique feature of the MO effect is non-reciprocity. The optical properties of non-reciprocal devices are different for two opposite directions of light propagation. The non-reciprocal effect can occur only in a MO material and the important optical non-reciprocal devices such as an optical isolator and an optical circulator can only be fabricated by utilizing the MO materials.

The MO effect is known to occur in a configuration when light propagates along a magnetic field. When the magnetic field is applied to a material, the electrons with spin directed along and opposite to the magnetic field have different energies. Since the electrons of one spin direction interact with light either of left or right circular polarization[5-7], light of the left and right circular polarizations experiences different refraction and absorption. When light is transmitted through a MO material, there is a difference of refractive indexes (Faraday effect) and optical absorption (magnetic circular dichroism (MCD effect)) for left and right circularly polarized light. When light is reflected from a MO material, the reflectivity of the right and left circular polarized light is different (polar and longitudinal Kerr effects). All above-mentioned effects have the same origin and the similar properties and will be referred as longitudinal MO effects.

It should be no MO effect in the case of the magnetic field applied perpendicularly to the light propagation direction. Since an electromagnetic wave is transverse, its polarization should be in a plane, which is perpendicular to the light propagation direction. The polarization rotation is possible only around an axis, which is out of this plane. It implies that in the case of transverse magnetization, the polarization rotation around axis parallel to the magnetic field is impossible and the interaction of photons with electrons of opposite spins does not depend on the polarization of light. Therefore, in this geometry the probability to excite electrons with opposite spins is equal for left and right circularly polarized light and light should not be sensitive to the energy difference of electrons with opposite spins. It should be no MO effect in this case. It is true in the case of free-space propagation. Light only experiences birefringence, which is proportional to the square intensity of the magnetic field. It is called the Cotton–Mouton effect (in the case of gases it is called the Voigt effect). It is a reciprocal effect and may be simply considered as a magnetically-induced anisotropy.

However, there are several cases when the non-reciprocal MO (nMO) effect, which is linearly proportional to the intensity of the magnetic field, was experimentally observed in a media magnetized perpendicular to the light propagation direction. For example, when light is reflected from a ferromagnetic metal and the magnetization of the metal is perpendicular to the incidence plane, the absorption of light by the metal is different for the two opposite magnetization directions (transverse Kerr effect). The second example is light propagation in an optical waveguide covered by a ferromagnetic metal. In the case of a magnetic field applied perpendicularly to the light propagation direction in a waveguide plane, the absorption of the waveguide mode is significantly different for two opposite directions of the magnetic field[8-12]. The third example is a waveguide covered by a transparent MO material, in this case the propagation speed of waveguide mode changes when the applied magnetic field is reversed[13,14]. In all above-mentioned cases the MO effect is non-reciprocal and linearly proportional to the intensity of the magnetic field[15]. The properties of the observed transverse nMO effect are different from those of the longitudinal nMO effect. Firstly, the best-known feature of the longitudinal nMO effect, the rotation of the polarization plane is not a feature of the transverse nMO effect. The transverse nMO effect does not cause any polarization rotation. Due to the transverse nMO effect the absorption and refractivity of light changes when the magnetic field is reversed. Secondly, the longitudinal nMO effect influences both orthogonal polarizations. The absorption and refraction of both left and right circularly polarized waves change when the magnetic field is reversed. By contrast, only one polarization is affected by transverse nMO effect and the orthogonal polarization is not sensitive to MO. Thirdly, in contrast to the longitudinal nMO effect the transverse nMO effect occurs only in layered structures. It has never been observed in free-space. Fourthly, the magnitude of the longitudinal nMO effect decreases in the case of light propagation along a multilayer structure, because of "TE-TM mode mismatch" problem[16]. In order to achieve high magnitude of the longitudinal MO effect in most of cases the interfaces should be graded[16]. By contrast, to achieve a substantial transverse nMO effect sharp interfaces are required[9].

Despite a long history of experimental and theoretical study of the transverse nMO effect, several important issues still remain unresolved. The origin of the transverse MO effect is unexplained. The general conditions, at which the effect can occur, are unclear. The factors influencing the magnitude of the effect are not determined. The reason for the difference in the properties of longitudinal and transverse nMO effects is not explained.

In this Letter we will explain the origin of the transverse nMO effect and we will describe the conditions, at which light may experience the transverse nMO effect. We will demonstrate that in contrast to the longitudinal nMO effect the transverse nMO effect has two contributions and by optimizing these contributions the effect may be significantly magnified. We will demonstrate that in the case of the surface plasmons, the magneto-optical figure-of-merit may increase from a few percents to above 100%. We will derive the scalar dispersion relation, which describes the transverse MO effect in cases of waveguide modes and surface plasmons propagating in a multilayer MO slab.

It should be noticed that in all cases the magnitude of the transverse nMO effect can be calculated by solving Maxwell's equations and utilizing an asymmetric permittivity tensor for the MO layer. It was calculated for transparent waveguides[13,14], hybrid waveguides[8,9] and surface plasmons[17-19]. Always, there was a good correspondence between calculated and experimental data. The correct description of both the longitudinal and the transverse nMO effects by the same permittivity tensor implies that the origins of the longitudinal and transverse nMO effects should be similar. The primary reason for both nMO effects is a splitting of energy bands for electrons with opposite spins in a magnetic field. However, the conditions, at which light is sensitive to that splitting, are different for the longitudinal and transverse nMO effects. In the case of the magnetic field perpendicular to the light propagation direction, these conditions are very specific and they mainly determine the properties of the transverse nMO effect. In the following we will describe the origin of the transverse nMO effect and the conditions at which the effect occurs.

Let us consider light propagating along the z-axis in a MO media and the magnetic field applied along the y-axis. In this case the electrons with opposite spins along the y-axis have different energy. As was mentioned above, in the case when light is polarized perpendicularly to its propagation direction in xy-plane, the probabilities to excite electrons with the opposite spins are equal and light will not experience any nMO effect. In order for the nMO effect to occur, the polarization of light should be rotating around the axis, which is along the magnetic field and perpendicular to the light propagation direction. Only in this case light interacts differently with electrons of opposite spins. The fact that the polarization rotation should be around axis perpendicular to the propagation direction means that light should have a polarization component along the propagation direction. Since an electromagnetic wave is transverse, it seems that such a polarization is forbidden. However, there is one special case when such a polarization is possible. It is the case when a wave has an evanescent component along the direction perpendicular to the wave propagation direction. For example, it is the case when the x-component of the wave vector has only an imaginary part and the z-component has only a real part. So the wave is described as

$$\vec{E}, \vec{H} \sim e^{\frac{2\pi}{\lambda} i \cdot (-c \cdot t + k_z z + i \cdot k_x x)} + c.c. \qquad (1)$$

The polarization of this wave may be either along the y-axis ($E_x, E_z = 0, E_y \neq 0$) or in the xz plane ($E_x, E_z \neq 0, E_y = 0$). In the case of polarization in the xz-plane, the condition for the wave to be transverse is

$$\vec{E} \cdot \vec{k} = E_z \cdot k_z - i \cdot k_x \cdot E_x = 0$$

or

$$\frac{E_z}{E_x} = i \cdot \frac{k_x}{k_z} \qquad (2)$$

If $\frac{k_x}{k_z} = 1$, the polarization is circular and the axis of polarization rotation is perpendicular to the wave propagation direction and along the magnetic field. This wave will interact only with electrons of one spin direction and will experience the transverse nMO effect. The wave (2) is transverse, but it has a polarization component along the propagation direction. The seeming contradiction in the above-mentioned properties is explained as follows. As required for a transverse wave, the wave polarization is perpendicular to the direction of the wave vector $\vec{k}$, but the $\vec{k}$ direction is not the propagation direction for this wave. For an electromagnetic wave the propagation direction should be determined by a direction of electromagnetic energy flow[7], which is defined by time-averaged Poynting vector $\vec{S} = \frac{1}{2}\text{Re}(\vec{E} \times \vec{H}^*)$ and for the wave (2) it is along z-direction[20], along which the wave has a polarization component.

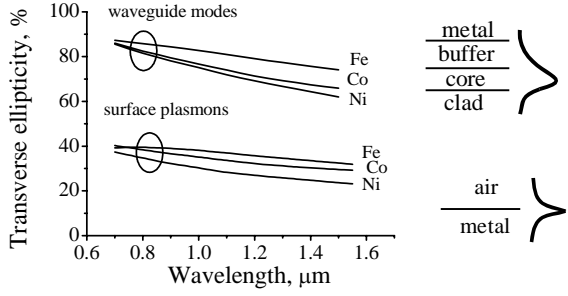

FIG.1. Transverse ellipticity of waveguide mode propagating in metal/AlGaAs waveguide and surface plasmon propagating at metal/air interface. 100% corresponds to a circular polarization. The upper and lower insets show the optical field distribution of a waveguide mode and a surface plasmon, respectively.

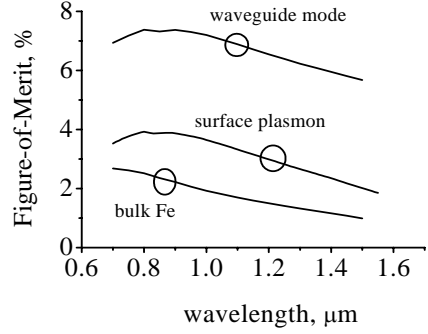

FIG.2. MO Figure-of-Merit (FoM) of the longitudinal nMO effect in Fe and FoM of transverse nMO effect for waveguide mode propagating in Fe/AlGaAs waveguide and for surface plasmons propagating at Fe/air interface.

From the above discussion it can be concluded that only a wave, which has an evanescent component and is elliptically polarized in a plane of wave propagation, may experience the transverse nMO effect. That condition explains several properties of the transverse nMO effect. The wave may have an evanescent component only in the vicinity of the interface. This explains why the transverse nMO effect occurs only in layered structures. Also, only one of two orthogonal polarization of wave (2) may experience the transverse MO effect. The polarization ($E_x, E_z = 0, E_y \neq 0$) is not rotating and light of such polarization will not experience any nMO effect.

The transverse ellipticity is essential for light to experience the transverse nMO effect. The transverse ellipticity $\chi$ is calculated as

$$\chi = 2 \cdot \text{Im}\left(\frac{E_z}{E_x}\right) \bigg/ \left(1 + \left|\frac{E_z}{E_x}\right|^2\right) \quad (3)$$

In the case of the circular polarization the transverse nMO effect will be strongest. Figure 1 shows calculated transverse ellipticity in cases of a waveguide mode propagating in Fe/AlGaAs, Co/AlGaAs, Ni/AlGaAs waveguides (buffer $Al_{0.5}Ga_{0.5}As$(300 nm)/core $Al_{0.3}Ga_{0.7}As$(1200 nm)/ clad $Al_{0.5}Ga_{0.5}As$) and in cases of a surface plasmon propagating along Fe/air, Co/air, Ni/air interfaces. In all cases the traverse ellipticity is significant. The ellipticity is close to circular for the waveguiding mode. That is the reason why the experimentally observed transverse nMO effect in waveguides is substantial[9-12]. The data were calculated by solving Maxwell's equations and utilizing known optical and magneto-optical constants of transition metals[19,20] and semiconductors[21].

It is interesting to compare the strength of the longitudinal and transverse nMO effects. Figure 2 shows the calculated MO Figure of Merit (FoM) for the longitudinal nMO effect in Fe, FoM of transverse nMO effect for a plasmon propagating at Fe/air interface and FoM of the transverse nMO effect for a waveguide mode propagating in Fe/$Al_{0.5}Ga_{0.5}As$ (300 nm)/$Al_{0.3}Ga_{0.7}As$ (1200 nm)/ $Al_{0.5}Ga_{0.5}As$ waveguides. FoM is a ratio of the magnetization dependent optical loss to the average loss. The nMO effect in the case of the transverse magnetization is 2-3 times stronger than in the case of the longitudinal magnetization.

In the following we will describe two major contributions to the transverse nMO effect. Since the effect may occur only for a wave having an evanescent component, the effect is pronounced near an interface. The first contribution comes from a bulk of MO material and the second contribution comes from the interface. To demonstrate this, let's consider a plain wave (1) propagating along a ferromagnetic-metal/dielectric interface. The magnetization of the metal is perpendicular to the light propagation direction. Light is absorbed by the metal and the flux of light energy dissipates along the propagation direction. Utilizing Poynting theorem[7], the optical loss is calculated as

$$loss = \frac{2\pi}{\lambda}\text{Im}(k_z) = -\frac{1}{2}\frac{\int \frac{\partial S_z}{\partial z}dx \cdot dy}{\int S_z \cdot dx \cdot dy} = \frac{1}{2}\frac{\int \frac{\partial U}{\partial t}dx \cdot dy}{\int S_z \cdot dx \cdot dy} = \frac{1}{2}\frac{\int_{metal} \frac{2\pi c}{\lambda}\frac{i}{8}(\varepsilon_{ik} - \varepsilon_{ki}^*)E_i E_k^* dx \cdot dy}{\int_{metal+dielectric} S_z \cdot dx \cdot dy} = \frac{2\pi c}{\lambda}\frac{1}{8}\left[\text{Im}(\varepsilon_0) + \text{Im}(\gamma) \cdot \chi\right]\frac{\int_{metal} |\vec{E}|^2 dx \cdot dy}{\int_{metal+dielectric} S_z \cdot dx \cdot dy} \quad (4)$$

where $\hat{\varepsilon} = \begin{pmatrix} \varepsilon_0 & 0 & -i\cdot\gamma \\ 0 & \varepsilon_0 & 0 \\ i\cdot\gamma & 0 & \varepsilon_0 \end{pmatrix}$ (5)

is the permittivity tensor of the ferromagnetic metal, $\vec{S}$ is the Poynting vector and $\chi$ is the transverse ellipticity (3).

When the magnetic field is reversed, the absorption of light changes, because of the transverse nMO effect. If we assume that the distribution of an optical field does not depend on the magnetic field, the ratio of integrals at the left part of (4) would be the same for both directions. That assumption means that the redistribution of the optical field due to the influence of the interface is ignored. Than, FoM of the bulk contribution to transverse nMO effect is derived from (4) as

$$FoM_{bulk} = 2\chi \frac{\mathrm{Im}(\gamma)}{\mathrm{Im}(\varepsilon_0)}$$ (6)

The FoM of the bulk contribution is linearly proportional to the transverse ellipticity. It supports given-above explanation of the origin of the transverse MO effect. It should be considered that optical field distribution near the interface depends on the refractive index of both the metal and the dielectric. In the case of non-zero transverse ellipticity the effective refractive index of the metal changes with reversal of the magnetic field due to MO effect. That leads to the redistribution of the optical field near the interface and the ratio of integrals at the left part of (4) will be different for opposite directions of the magnetic field. That defines the interface contribution to the transverse nMO effect. The interface contribution is explained as follows. The MO change of the effective refractive index of the metal redistributes relative amounts of light energy between the dielectric and the metal. Since the absorption by the metal is proportional to the amount of light inside it and this amount changes with reversal of the magnetic field, the optical loss of the wave will also change with the reversal of the magnetic field. For the transverse nMO effect, both the bulk and interface contributions are substantial, often have opposite signs and should be carefully considered. For example, it was observed experimentally[8] that in the case of Co:AlGaAs waveguide the sign of the transverse nMO effect changes when an AlGaAs buffer layer between waveguide core and cobalt is replaced with a $SiO_2$ buffer layer. This is because the bulk contribution is the same in both cases, but interface contribution is of opposite signs and it is negligible in the case of Co:AlGaAs interface and is about two times greater than the bulk contribution in the case of a $Co:SiO_2$ interface. Another example, which demonstrates the importance of the interface contribution, is the surface plasmons propagation at a Fe:MgO:AlGaAs interface. Figure 3 shows calculated plasmon's 1/e propagation distance and MO FoM as a function of MgO thickness. The structure does not support the plasmons when MgO thickness is between 14 nm and 110 nm. Near "cutoff" thicknesses of 14 nm and 110 nm, the MO FoM significantly increases and the plasmon's propagation distance becomes longer. It is explained as follows. When MgO thickness approaches a "cutoff" thickness, the penetration depth of plasmon's optical field into AlGaAs increases significantly, but the penetration depth into the iron remains unchanged. Therefore, the optical field is "pushed out" from the metal. That causes a significant decrease of optical loss and increase of the plasmon's propagation distance. A change of the refractive index of the metal changes the value of the "cutoff" thickness. The plasmon's optical loss sharply decreases with the MgO thickness approaching the "cutoff" thickness. Therefore, in the case of the MgO thickness close to the "cutoff" thickness, a small MO change of the refractivity of the metal causes a significant MO change of plasmon's optical loss. That causes a substantial transverse nMO effect. As seen from Fig.3, the MO FoM may reach 100%. This is a significant enhancement of nMO effect compared to 2-3% of the MO FoM, which is the most that can be achieved for the longitudinal nMO effect in Fe (Fig.2).

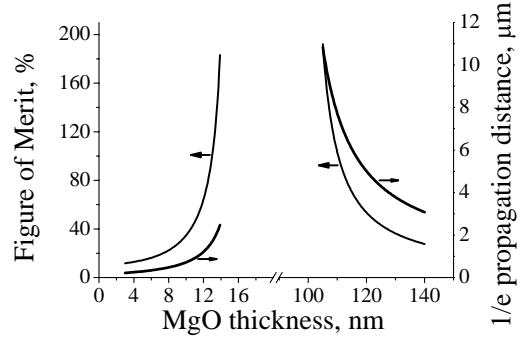

FIG.3. FoM and 1/e propagation distance for surface plasmons propagating at $Fe/MgO/Al_{0.5}Ga_{0.5}As$ interface ($\lambda$=800 nm).

Next, we will derive the dispersion relation, which describes the transverse MO effect in cases of waveguide modes and surface plasmons propagating in a multilayer MO slab. Since the permittivity tensor of a MO material has non-zero off-diagonal components, the wave propagation in MO structure is conventionally described by a dispersion relation, which is a combination of (4x4) matrixes[19,26]. Because of the complexity of this dispersion relation, the approximation of small off-diagonal components is often utilized[26]. However, in the case of the transition metals, the off-diagonal components have the same order of magnitude as the diagonal components and that approximation is not always valid. In the following, without use of any approximations we will derive a scalar dispersion relation, which described the transverse nMO effect in a multilayer MO slab. The availability of the simple scalar dispersion relation significantly simplifies analysis and predictions for the transverse nMO effect.

Let's us consider a multilayer MO slab, where $\hat{\varepsilon}_j$ and $t_j$ the permittivity tensor (5) and the thickness of each j-layer, respectively. The layers of the slab are infinite in the xy-plane and the wave propagation direction is along z-direction. For a plain wave

$$E_x, E_z, H_y \sim e^{\frac{2\pi}{\lambda}i(-c\cdot t + k_x x + k_z z)} \quad (7)$$

a solution for Maxwell's equations:

$$rot(\vec{H}) = \frac{1}{c}\frac{\partial}{\partial t}(\hat{\varepsilon}\cdot\vec{E}); \quad rot(\vec{E}) = -\frac{1}{c}\frac{\partial}{\partial t}(\vec{H}) \quad (8)$$

will be $k_x^2 = \varepsilon_0 - k_z^2 - \gamma^2/\varepsilon_0$

$$\begin{bmatrix} E_x \\ H_y \end{bmatrix} = \begin{bmatrix} \dfrac{i\cdot\gamma - k_z k_x}{\varepsilon_0 - k_z^2} \\ \dfrac{i\cdot\gamma\cdot k_z - \varepsilon_0\cdot k_x}{\varepsilon_0 - k_z^2} \end{bmatrix} E_z \quad (9)$$

The optical field in j-layer can be described as

$$\begin{bmatrix} E_z^j \\ H_y^j \end{bmatrix} = \left\{ A_f^j e^{i\frac{2\pi}{\lambda}k_{xj}(x-t_j)}\begin{bmatrix} 1 \\ \dfrac{i\cdot\gamma_j\cdot k_z - \varepsilon_{0j}\cdot k_{xj}}{\varepsilon_{0j} - k_z^2} \end{bmatrix} + A_b^j e^{-i\frac{2\pi}{\lambda}k_{xj}(x-t_j)}\begin{bmatrix} 1 \\ \dfrac{i\cdot\gamma_j\cdot k_z + \varepsilon_{0j}\cdot k_{xj}}{\varepsilon_{0j} - k_z^2} \end{bmatrix} \right\} e^{\frac{2\pi}{\lambda}i(-c\cdot t + k_z z)} + c.c. \quad (10)$$

where $A_f^j$ and $A_b^j$ are unknowns and c.c. is complex conjugate. Introducing new unknowns

$$Z_j = \frac{A_f^j - A_b^j}{A_f^j + A_b^j} \quad B_j = A_f^j + A_b^j \quad (11)$$

and applying boundary conditions $E_z = const$ $H_y = const$ at boundary between j and j+1 layers, we have

$$Z_j = T_j\left[F_{j,j+1}\left[Z_{j+1}\right]\right] \quad (12)$$

where
$$T_j[Z] = \frac{Z - i\cdot tan\left(\frac{2\pi}{\lambda}k_{xj}t_j\right)}{1 - i\cdot Z\cdot tan\left(\frac{2\pi}{\lambda}k_{xj}t_j\right)}$$

$$F_{j,j+1}[Z] = \left[i\cdot\gamma_j\cdot k_z - \left(i\cdot\gamma_{j+1}\cdot k_z - Z\cdot\varepsilon_{0(j+1)}\cdot k_{x(j+1)}\right)\frac{\varepsilon_{0j} - k_z^2}{\varepsilon_{0(j+1)} - k_z^2}\right]\Big/\left(\varepsilon_{0j}\cdot k_{xj}\right)$$

In the case of a evanescent wave, $|\vec{E}|\to 0$ when $x\to\infty$, which leads to $A_b^n = 0$ $Z_n = 1$. The general solution describing transverse nMO effect in the case of a plain wave propagating in a MO multilayer slab will be

$$Z_j = T_j\left[F_{j,j+1}\left[...\left[T_j\left[F_{j,j+1}\left[...F^{n,n-1}[1]\right]\right]\right]\right]\right] \quad (13)$$

The Eqn. (13) can be used to derive reflectivity of MO multilayer and dispersion relation for surface plasmons and waveguide modes. If the plain wave propagates in layer 1 and is reflected by a MO multilayer, the reflection coefficient will be

$$R = \frac{A_{bi}}{A_{fi}} = \frac{1-Z_1}{1+Z_1} \quad (14)$$

where

$$Z_1 = F_{1,2}\left[...\left[T_j\left[F_{j,j-1}\left[...F^{n,n-1}[1]\right]\right]\right]\right] \quad (15)$$

where n is the number of MO layers. The Eqns. (14) and (15) describe transverse Kerr effect in the case of the reflection from the multilayer.

In case of surface plasmon, light is confined near the metal/dielectric interface, so $|\vec{E}|\to 0$ when $x\to\pm\infty$. The dispersion relation for surface plasmons will be

$$F_{1,2}\left[...\left[T_i\left[F_{i,i-1}\left[...F^{n,n-1}[1]\right]\right]\right]\right] = -1. \quad (16)$$

Next let's consider an optical waveguide, in which layer 1 is a core, layer 2 is a cladding layer under the core and the MO multilayer on the top covering the core. The dispersion relation for the waveguide mode can be derived as

$$k_{x1}d_1 = a\tan(i\cdot Z_1) + a\tan\left(i\frac{\varepsilon_{02}}{\varepsilon_1}\frac{k_{x1}}{k_{x2}}\right) + m\cdot\pi \quad (17)$$

where $Z_1$ determined by (15) and m is a mode number.

Another interesting effect can be predicted from understanding of the origin of the transverse nMO effect. Recently, several new designs of photonic devices have been proposed, in which spin-polarized electrons are excited by circular-polarized light[24,25]. However, it is difficult to use optical waveguides in such devices, because of the difficulties in achieving circular polarization of a waveguiding light due to "TE-TM mode mismatch problem". By contrast, utilizing the transverse MO effect, the spin-polarized electrons may be efficiently excited in waveguiding devices. For example, let us consider an optical waveguide, in which the transverse magnetic mode (TM mode) propagates. If the waveguide contains a semiconductor layer and photon energy of light is slightly above the semiconductor bandgap, light will experience absorption and will excite electrons in this layer. As was explained above, if the optical field in this layer has an evanescent component, the polarization of light will be transverse elliptical. Therefore, the photo-exited electrons will be spin-polarized and the direction of the spin will be perpendicular to the light propagation direction. This effect does not require any mode phase matching and it could be utilized for efficient and reliable excitation of spin polarized electrons in future spintronics and spin-photonics devices.

In conclusion, the origin and the properties of the transverse nMO effect have been investigated. The transverse nMO effect occurs because the polarization of light, which has an evanescent component, rotates around an axis, which is parallel to the magnetic field and perpendicular to the light propagation direction. The transverse nMO effect is comparable to and often greater than the longitudinal nMO effect. In contrast to the longitudinal nMO effect, the transverse nMO can be significantly magnified by optimizing the device structure. We have demonstrated that in the case of surface plasmons the magneto-optical figure-of-merit may increase from few percents to above 100%. Because of this unique property, the transverse nMO effect has an advantage for usage in a variety of applications where only the longitudinal nMO effect is currently utilized.